\documentclass{aastex} 
\usepackage{spr-astr-addons}

\begin{document}

\title{Empirical Constraints for the Magnitude and Composition of Galactic Winds} 
\shorttitle{Magnitude and Composition of Galactic Winds}
\shortauthors{Zahid et al.}

\author{ H. Jabran Zahid\altaffilmark{1}}
\affil{University of Hawaii at Manoa, Institute for Astronomy} %, 2680 Woodlawn Dr., Honolulu,  HI 96822, USA\\
\altaffiltext{1}{Smithsonian Astrophysical Observatory Predoctoral Fellow}
\and 
\author{Paul Torrey} 
\affil{Harvard-Smithsonian Center for Astrophysics} %, 60 Garden Street, Cambridge, MA, 02138, USA\\ 
\and 
\author{Mark Vogelsberger\altaffilmark{2}}
\affil{Harvard-Smithsonian Center for Astrophysics} %, 60 Garden Street, Cambridge, MA, 02138, USA\\ 
\altaffiltext{2}{Hubble Fellow}
\author{Lars Hernquist}
\affil{Harvard-Smithsonian Center for Astrophysics} %, 60 Garden Street, Cambridge, MA, 02138, USA\\ 
\and 
\author{Lisa Kewley}
\affil{Australian National University, Research School of Astronomy and Astrophysics}
\and 
\author{Romeel Dav\'e}
\affil{Astronomy Department, University of Arizona} %, Tucson, AZ 85721, USA\\
\affil{University of the Western Cape, Bellville} %, Capte Town 7535, South Africa\\
\affil{South African Astronomical Observatories} %, Observatory, Cape Town 7925, South Africa\\
\affil{African Institute for Mathematical Sciences}

%\maketitle
\begin{abstract}

Galactic winds are a key physical mechanism for understanding galaxy
formation and evolution, yet empirical and theoretical constraints for
the character of winds are limited and discrepant. Recent empirical
models find that local star-forming galaxies have a deficit of oxygen
that scales with galaxy stellar mass. The oxygen deficit provides
unique empirical constraints on the magnitude of mass loss, composition
of outflowing material and metal reaccretion onto galaxies. We
formulate the oxygen deficit constraints so they may be easily
implemented into theoretical models of galaxy evolution. We
parameterize an effective metal loading factor which combines the
uncertainties of metal outflows and metal reaccretion into a single
function of galaxy virial velocity. We determine the effective metal
loading factor by forward-fitting the oxygen deficit. The effective
metal loading factor we derive has important implications for the
implementation of mass loss in models of galaxy evolution.

\end{abstract}


\section{Introduction}

Galaxy scale winds are fundamental to galaxy evolution.  The observed
baryon content of galaxies is substantially below the cosmic baryon
fraction \citep{Papastergis2012}.  To account for this deficit, galaxy
formation theories require mechanisms to reduce the efficiency with
which galaxies grow \citep[e.g.,][]{Springel2003a}.  Consequently,
strong feedback which is capable of launching galactic scale outflows
is central to semi-analytic and hydrodynamical galaxy formation
models~\citep[e.g.,][ plus many others]{Somerville1999, Springel2003a,
Schaye2010, Dave2011a, Vogelsberger2013}. In
simulations, energy and/or momentum injected by massive stars is
capable of driving gas out of galaxies.  Although these outflows are
primarily required to regulate the growth of galaxies, they also drive
metals out of the interstellar medium (ISM).  Outflows reduce the
metal content in galaxies and contribute to the enrichment of the
circumgalactic and intergalactic medium \citep[e.g.][]{Springel2003a,
Oppenheimer2006, Dave2011a}.

Despite the need for outflows to regulate the growth of galaxies, the physical properties of galactic scale winds are poorly determined observationally. The lack of understanding is partly due to the complex multi-phase structure of the gas, which can only be characterized in detail with observations over a broad range of wavelengths \citep{Veilleux2005}. While absorption line studies permit direct measurements of outflow velocities \citep[e.g.][]{Heckman2000, Rupke2005, Weiner2009, Chen2010, Erb2012, Martin2012, Rubin2013}, estimates of the magnitude of mass loss in winds are much more difficult to obtain. Moreover, the metallicity of the wind material is constrained by only a few observations~\citep[e.g.,][]{Martin2002}.

In the absence of a complete understanding of the character of
outflows, theories employ wind prescriptions that are tuned to
reproduce the observed properties of the galaxy population (e.g., the
galaxy stellar mass function).  This is typically achieved by assuming
that a constant amount of energy or momentum is injected per unit star
formation and that the wind speed scales in proportion to galaxy
escape velocity~\citep{Martin2005}.  However, in detail, the
normalization and scaling of outflows vary significantly between
implementations.  For example, the galaxy stellar mass functions
derived by~\citet{Dave2011a} and~\citet{Puchwein2013} both provide
satisfactory fits to the stellar mass function of local galaxies; this
is in spite of mass loading factors that have normalization offsets of
an order of magnitude for low stellar mass (i.e. $M_*\sim 10^9
M_\odot$) systems.  The uncertainties in our understanding of the
physical properties of galactic winds are further complicated by
numerical approximations that vary substantially from one simulation
code to another \citep[e.g][]{Vogelsberger2012, Keres2012,
Sijacki2012, Torrey2012, Nelson2013}.

Gas outflows deplete the heavy element content of galaxies.  Typically it is assumed that the wind material has the same metallicity as the ambient ISM. However, the actual wind metallicity relative to the ISM may be greater if it is primarily comprised of supernova ejecta or
substantially depressed if a sufficient amount of metal-poor gas is entrained as the wind propagates out of the galaxy.  The total amount of metals ejected from the ISM will be proportional to the magnitude and metallicity of the outflowing gas, modulo the amount of metals ejected and subsequently reaccreted.  In the absence of direct measurements characterizing these physical processes, we must rely on empirical constraints for the total metal loss in the local galaxy population to infer the properties of outflows and reaccretion.

We provide empirical constraints for outflows  that can be easily implemented in galaxy formation models. In Section 2 we review the empirical oxygen mass loss estimates of~\citet{Zahid2012b} and the formalism for outflows typically employed  in galaxy formation models. In Section 3 we introduce an effective metal loading factor which describes the loss of heavy elements from galaxies under the influence of both galactic outflows and inflows and derive the best fit for the effective metal loading factor by forward-fitting the empirical oxygen mass loss estimates. In Section 4 we directly compare our fit with parameterizations of outflows currently used in galaxy formation models and summarize our results in Section 5. We adopt a standard cosmology $(H_{0}, \Omega_{m}, \Omega_{\Lambda}) = (70$ km s$^{-1}$ Mpc$^{-1}$, 0.3, 0.7 and a \citet{Chabrier2003} initial mass function (IMF).

\section{Formalism}
\subsection{The Oxygen Budget}

\begin{figure}
\begin{center}
\includegraphics[width=\columnwidth]{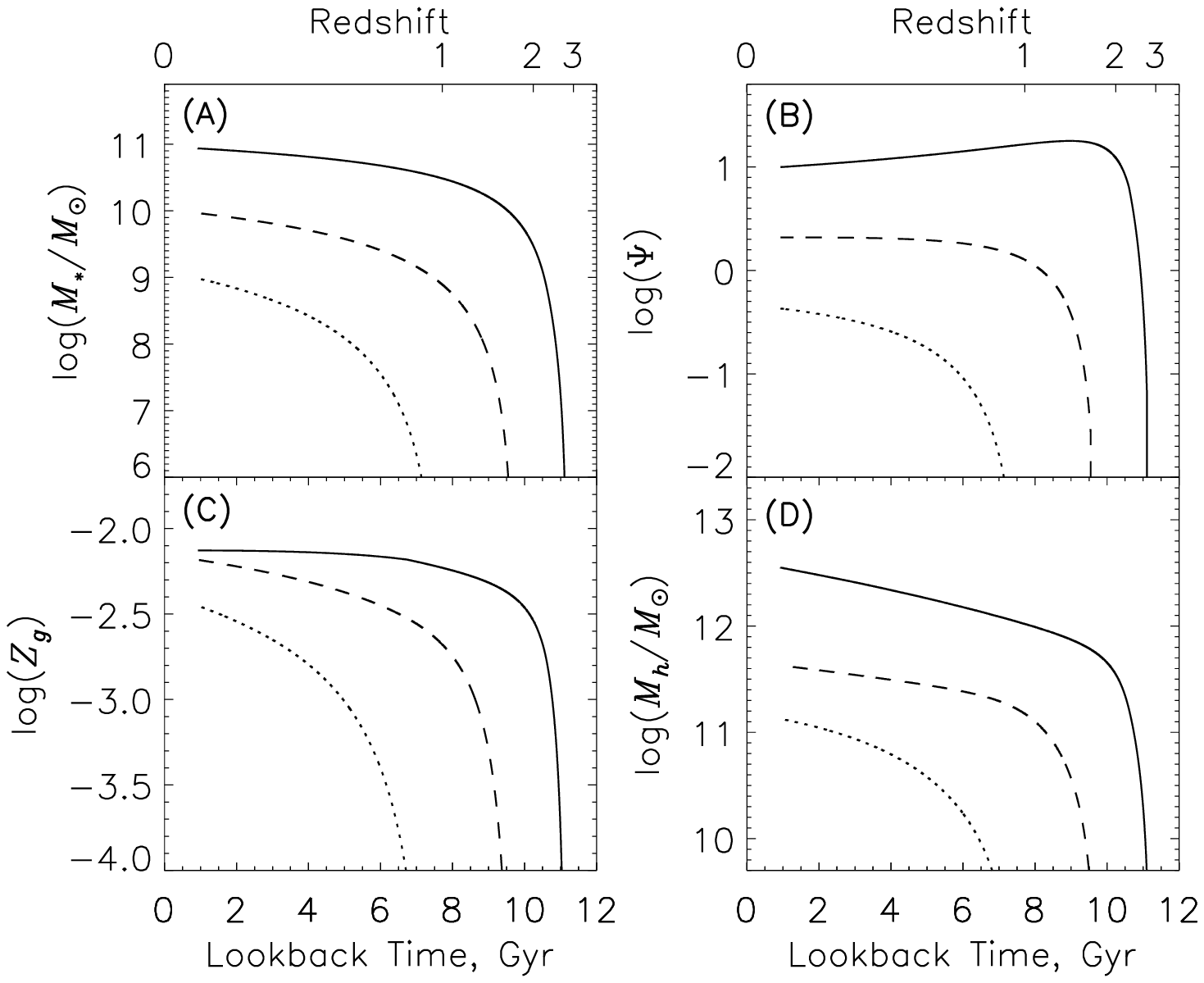}
\end{center}
\caption{Panels (A) and (B): Stellar mass and star formation history tracks for three 
example model star-forming galaxies as required by the multi-epoch observations of the MS relation, respectively. Panel (C): Chemical history tracks for the model galaxies determined from multi-epoch observations of the MZ relation. Panel (D): Halo mass tracks inferred from the inversion of the stellar mass-halo mass function of \citet{Behroozi2012}.}
\label{fig:model}
\end{figure}

Estimates for the total oxygen budget of local star-forming galaxies is provided by \citet{Zahid2012b}. We briefly summarize the formalism here. The oxygen balance can be characterized as:
\begin{equation}
M_T^o = M_g^o + M_\ast^o + \Delta M^o ,
\label{eq:oxy_inventory}
\end{equation}
where $M_T^o$ is the total oxygen mass produced by stellar evolution, $M_g^o$ is the mass of oxygen currently in the ISM, $M_\ast^o$ is the mass of oxygen locked up in stars, and $\Delta M^o$ is the remaining unaccounted for oxygen which resides in the intergalactic medium (IGM) and/or circumgalactic medium (CGM).  Three of the four terms in this equation can be empirically constrained.

The total mass of oxygen produced is $M_T^o = y M_\ast/(1-R)$. Here, $y$ is the total mass of oxygen formed per unit mass of star formation and the factor $M_\ast/(1-R)$ is the total integrated stellar mass. This is the current mass in stars, $M_\ast$, increased by a factor $1/(1-R)$ to account for gas recycling. Here $R$ is fraction of mass that forms into stars and is later returned back to the ISM by stellar winds and supernovae. This fraction can range between 0.2 - 0.5 depending on the choice of IMF \citep{Leitner2011, Zahid2012b}. In this work we adopt $R=0.35$. The mass of oxygen in the ISM is $M_g^o = Z_g M_g$, where $M_g$ is the galaxy gas mass and $Z_g$ is the gas-phase oxygen abundance. The total oxygen mass locked up in stars is
\begin{equation}
M_\ast^o = (1-R)\int Z_g(t) \; \Psi(t)  \; dt.
\label{eq:oxy_stellar}
\end{equation}
Here $\Psi(t)$ is the star formation rate (SFR) as a function of time. Evaluating Equation~\ref{eq:oxy_stellar} requires constraints on the chemical and star formation histories of the local star-forming galaxy population. Empirically, stellar mass growth and chemical evolution are constrained by the fact that galaxies evolve along a redshift dependent stellar mass-SFR (MS) relation, $\Psi = \Psi(z,M_\ast)$, and that galaxies evolve along a redshift-dependent mass-metallicity (MZ) relation, $Z_g = Z_g(z,M_*)$.

In~\citet{Zahid2012b}, multi-epoch observations are used to define the MS and MZ relation tracks which star-forming galaxies follow.  Figures \ref{fig:model}A and \ref{fig:model}B show the stellar mass and star formation history tracks of three model galaxies assuming evolution along the MS relation. Figure \ref{fig:model}C shows the gas-phase abundance tracks required by the observed MZ relation at several epochs and the stellar mass growth inferred from the MS relation. The  mass locked up in stars is estimated by evaluating Equation  \ref{eq:oxy_stellar}.  Using empirical estimates for $M_T^o$, $M_g^o$ and $M_\ast^o$ we can determine $\Delta M^o$ from Equation \ref{eq:oxy_inventory}. The oxygen deficit, $\Delta M^o$, represents the magnitude of oxygen mass loss that has occurred to ensure that local galaxies fall along the observed MZ relation.

Systematic uncertainties in the parameters required for conducting the oxygen census yield a range of possibilities for the oxygen deficit as a function of galaxy stellar mass \citep[see][]{Zahid2012b}.  We adopt the oxygen deficit that is consistent with independent
estimates of the oxygen content of hot halos of star-forming galaxies \citep{Tumlinson2011}. The empirically determined oxygen deficit
is
\begin{equation}
\Delta M^o (M_\ast) = 4.48\times10^{7}  \left( \frac{M_\ast}{10^{10} M_\odot} \right)^{1.13} \,\,[M_\odot].
\label{eq:od_fit}
\end{equation}
A robust conclusion of \citet{Zahid2012b} is that oxygen mass loss is more significant for massive galaxies. We note that the estimates of \citet{Tumlinson2011} are lower limits and the total oxygen mass in halos may be larger by a factor of a few \citep[J. Werk, private communication;][]{Werk2013}. This would lead to a commensurate increase in the derived oxygen deficit. 

The key step for parameterizing the oxygen loss for use in galaxy formation models is to cast the deficit in terms of quantities related to the galaxy potential well.  We determine the halo mass of galaxies in our model, $M_h = M_h(z, M_\ast)$, using the stellar mass - halo mass relation of \citet{Behroozi2012}\footnote{The relation is not directly invertible and we use tables provided by P. Behroozi (private communication).}. Our model of stellar mass growth taken together with the redshift dependent stellar mass - halo mass relation allows us to track the evolution of the galaxy potential well as a function of redshift. Figure \ref{fig:model}D shows the halo mass as a function of time for the three model galaxies. The evolving virial velocity of each galaxy is easily derived from the halo mass as a function of redshift and stellar mass and is given by Equation 19 in \citet[][and references therein]{Peeples2011}. In Section 3, we model the effective metal loading factor as a function of virial velocity and forward-fit this model to the observed oxygen deficit.

\subsection{Ejecting Oxygen from Galaxies}

Our primary goal in this contribution is to provide the new empirical
constraints on galaxy mass loss from~\citet{Zahid2012b} in a form that
is directly applicable in galaxy formation models. The oxygen mass
loss can be accounted for by enriched material that is ejected
from the galaxy into the CGM and/or the IGM.  In either case, the
oxygen mass loss is caused by galactic outflows.  We note that most
semi-analytic models and hydrodynamic simulations of galaxy formation
rely on star formation driven winds as the primary (if not only) mass
loss mechanism that operates in low-mass galaxies ($M_\ast \lesssim
10^{11}M_\odot, M_h \lesssim 10^{12} M_\odot$).  These winds originate
from deep within the galactic potential wells and are a
form of ``ejective" feedback. They regulate the growth of galaxies and are also thought to be primary mechanism for metal loss in
star-forming galaxies.  While black hole growth may also drive winds \citep[e.g.][]{Rupke2011}, we do not consider black hole driven feedback here since it is likely to be prevalent in only the most massive galaxies \citep[e.g.][]{Hopkins2006, Hopkins2008, Somerville2008}.

A general form of the oxygen deficit integrated over the lifetime of a galaxy is
\begin{equation}
\Delta M^o = \int \left[Z_w(t)\dot{M}_w(t) - Z_{acc}(t) \dot{M}_{acc}(t) \right]dt ,
\label{eq:int_oxy_def}
\end{equation}
where $Z_w$, $\dot{M}_w$, $Z_{acc}$ and $\dot{M}_{acc}$ are the wind metallicity, wind mass loss rate, accretion metallicity and accretion rate, respectively. None of the terms inside the integral on the right hand side of Equation~\ref{eq:int_oxy_def} are well constrained observationally. In the following section we present our primary result; a parameterization of the oxygen deficit that combines the uncertainties of each of the terms in Equation \ref{eq:int_oxy_def}.

Often in galaxy formation models two basic assumptions are made regarding galactic outflows. First, it is assumed that the mass loss rate from star-formation driven winds is directly proportional to the star formation rate. This relationship is parameterized in the mass loading factor, $\eta_w$, such that $\dot{M}_w = \eta_w \Psi$.  The mass loading factor is tuned so that models reproduce basic properties of the galaxy population.  Second, the wind metallicity is parameterized by a metal loading factor, $\gamma_w$, such that $Z_w = \gamma_w Z_{g}$. In uniform wind models it is assumed that the metallicity of the outflowing wind material is identical to that of the ambient ISM from which the wind is launched. The metal loading factor in many models of galaxy formation is set to unity  though there is no compelling observational evidence to suggest that this is the case.

\section{Effective Metal Loading Factor}

To take advantage of the empirical constraints on $\Delta M^o$ described in Section 2, we can rewrite the oxygen deficit as
\begin{equation}
\Delta M^o = \int \zeta(t)  \; \Psi(t) \; Z_g(t) dt
\label{eqn:start2}
\end{equation}
where
\begin{equation}
\zeta  = \eta_w \gamma_w - \eta_{acc} \gamma_{acc} 
\label{eqn:best_fit_metal_loading}
\end{equation}
is the effective metal loading factor.  In this form, the metallicity of the wind and accreted material are expressed in terms the ISM metallicity using the relative metallicity parameters $\gamma_w$ and $\gamma_{acc}$.  Similarly, the rate of mass ejection in the wind and the rate of mass accretion are expressed in terms of the star formation rate using the relative mass loading parameters $\eta_w$ and $\eta_{acc}$.  The advantage of the formulation presented in Equation \ref{eqn:start2} is that we have empirical constraints on the oxygen deficit, $\Delta M^o$, the metallicity, $Z_g$, and the star formation rate, $\Psi$, leaving only the effective metal loading factor undetermined.  Thus, by adopting  a functional form for the effective metal loading factor we can forward-fit  the oxygen deficit to derive the best fit parameters.  The key step taken here is combining our uncertainty about the mass loading, $\eta$, and relative metallicity, $\gamma$, into a single function which can be empirically constrained.

\begin{figure}
\begin{center}
\includegraphics[width=\columnwidth]{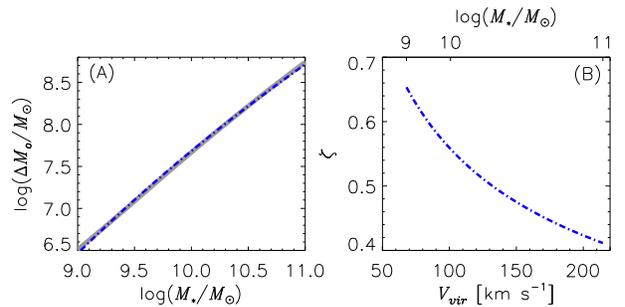}
\end{center}
\caption{ Panel (A): Forward-fit of the effective metal loading factor given by Equation \ref{eq:zeta}. The oxygen deficit (Equation \ref{eq:od_fit}) is shown by the solid grey curve.The best fit model is given by the dot-dashed blue curve.  Panel (B): Effective mass loading factor derived from panel (A) plotted as a function of virial velocity.}
\label{fig:model_result}
\end{figure}

A best fit effective metal loading factor can be derived based on Equation~\ref{eqn:start2} by comparing  to the empirically determined oxygen deficit (see Section 2.1). To facilitate the fitting procedure, we parameterize the effective 
metal loading factor as 
\begin{equation}
\zeta(V_{vir}) = \alpha \left( \frac{100\,\, \mathrm{km} \,\, \mathrm{s}^{-1}}{V_{vir}}\right)^\beta 
\label{eq:zeta}
\end{equation}
where $\alpha$ and $\beta$ set the normalization and slope, respectively.  The power law functional form is adopted because a similar functional form is used to parameterize wind mass loading factors in galaxy formation models.  We determine the best-fit parameter values by minimizing the residuals between the empirically derived oxygen deficit given by Equation \ref{eq:od_fit} and the oxygen deficit we calculate from integrating Equation \ref{eqn:start2} using the same empirical model of stellar mass growth, halo mass growth and chemical evolution as described in Section 2.1. The best fit parameters are $\alpha = 0.57$ and $\beta = 0.41$ and the best-fit result for the effective metal loading factor is shown by the blue dot-dashed curve in Figure \ref{fig:model_result}.

The effective metal loading factor we present can be straightforwardly applied as an empirical constraint for mass loss in simulations. Equation \ref{eqn:best_fit_metal_loading} can be solved for any factor(s) that is (are) not constrained within the simulation. If the best-fit effective metal loading factor parameterized by Equation \ref{eq:zeta} is adopted, the solution for any unconstrained factor(s) in Equation \ref{eqn:best_fit_metal_loading} will necessarily be consistent with the empirical mass loss constraints based on the oxygen deficit of \citet{Zahid2012b}. As empirical constraints for metal loss in galaxies improve, the approach presented here provides a straightforward methodology for implementing empirical constraints within theoretical models.

\begin{table}
\begin{center}
\caption{Mass loading factors found in the literature and derived in this work. For simplicity, when necessary we assume $v_w = 3V_{vir}$ where $v_w$ is the wind velocity.}
\label{table:mass_loading_factors}
\begin{tabular}{ l  l l l l  }
 Model & $\alpha$ & $\beta$ \\
\tableline
% Best Fit (this work)    & 0.57 & 0.41\\
 \citet{Peeples2011}  & 0.53 & $2$  \\
 \citet{Vogelsberger2013}     & 50 & $2$  \\
 \citet{Puchwein2013}    & 10 &  $2$ \\
 \citet{Okamoto2010}  & 13 &  $ 2 $ \\
 \citet{Dave2011a}  & 2.25 &  $1$ \\
 \citet{HopkinsWinds}  & 7 & $1.1$  \\
\tableline
\end{tabular}
\end{center}
\end{table}

\section{Discussion}

Constraining the cycling of gas and metals in and out of galaxies is key for understanding the chemical evolution of the universe. The properties of gas cycling are defined by four quantities; the wind and accretion mass loading  ($\eta_w$ and $\eta_{acc}$, respectively) and the relative metallicity of outflowing  and inflowing gas with respect to the ISM metallicity ($\gamma_w$ and $\gamma_{acc}$, respectively).  In the absence of direct observational constraints, the magnitude of the mass loading factors in models of galaxy formation are constrained by the baryonic mass function. The oxygen deficit, which is set by the cumulative effect of oxygen ejection and reaccretion, provides independent  constraints on the metallicity of outflows and inflows. However, the magnitude and composition of both outflows and inflows are only \emph{relatively} constrained by  the observations and therefore are subject to significant degeneracies. In consideration of these uncertainties, the effective metal loading factor we derive combines the uncertainties of the mass loading and composition of outflows with the efficiency of metal reaccretion into a single empirically constrained function of halo virial velocity. %The derived effective metal loading factor is of direct relevance to characterizing mass loss in cosmological simulations.

\begin{figure*}
\begin{center}
\includegraphics[width=2\columnwidth]{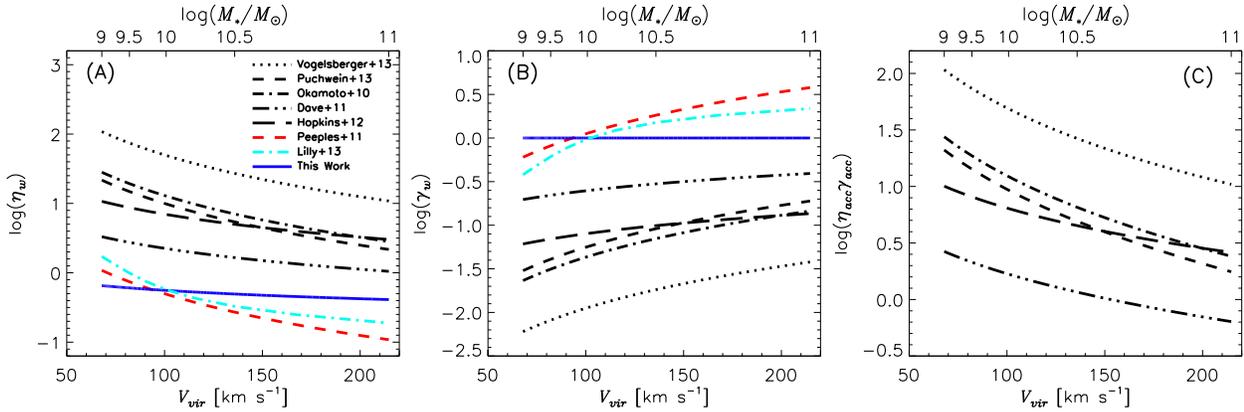}
\end{center}
\caption{(A) Comparison of various mass-loading factors compiled from the literature.  The mass loading factors that are implemented in various theoretical models are shown by the black curves. The empirically derived mass-loading factors are shown by the colored curves. Under the assumption of a uniform wind and negligible metal accretion, the metal loading factor we derive reduces to the mass loading factor. The solid blue curve is our metal loading factor assuming $\gamma_w = 1$ and $\eta_{acc}\gamma_{acc} << \eta_{w}\gamma_{w}$. (B) The metallicity of outflowing material for the various mass loading factors such that galaxies are consistent with the derived oxygen deficit and galaxies in theoretical models fall on the local MZ relation under the assumption that metal reaccretion is negligible. (C) The metal reaccretion efficiency assuming a uniform wind model ($\gamma_w = 1$) for various mass loading factors such that galaxies are consistent with the derived oxygen deficit and fall on the local MZ relation. Under these assumptions the metal reaccretion efficiencies of \citep{Peeples2011} and \citep{Lilly2013} are negative for massive galaxies which is unphysical and so are not plotted.}
\label{fig:eta}
\end{figure*}

%%%%%%%%%%%%%%%%%%%%%%
\begin{comment}
\begin{table}
\begin{center}
\label{table:reaccretion_efficiency}
\caption{Mass loading factors found in the literature and derived in this work. For simplicity, when necessary we assume $v_w = 3V_{vir}$ where $v_w$ is the wind velocity.}
\label{table:mass_loading_factors}
\begin{tabular}{ l  l l l l  }
 Model & $\mu$ & $\rho$ \\
\tableline
 \citet{Vogelsberger2013}  & 50 & $2$  \\
 \citet{Puchwein2013}    & 9 &  $2.1$ \\
 \citet{Okamoto2010}  & 12 &  $ 2.1 $ \\
 \citet{Dave2011a}  & 1.7 &  $1.2$ \\
 \citet{HopkinsWinds}  & 6 & $1.2$  \\
\tableline
\end{tabular}
\end{center}
\end{table}
\end{comment}
%%%%%%%%%%%%%%%%%%%%%%

Galaxies have baryonic content well below the simple prediction of the cosmological baryon fraction \citep[e.g.][]{Papastergis2012}. Thus, massive outflows are required in simulations in order to reproduce the observed distribution of stellar masses in galaxies \citep{Springel2003b}. Variable wind models, where mass loading scales with host galaxy properties, are successful at reproducing a broader range of observations \citep[e.g.][]{Oppenheimer2006, Finlator2008, Dave2011a, Dave2011b, HopkinsWinds, Puchwein2013}. In contrast, empirically derived mass loading factors are principally constrained by the metal content of galaxies \citep[e.g.][]{Peeples2011, Lilly2013}. Figure \ref{fig:eta}A shows mass loading factors found in the literature plotted as a function of galaxy virial velocity. Mass loading factors applied in theoretical models are indicated by the black curves \citep[i.e.][]{Okamoto2010, Dave2011a, HopkinsWinds, Puchwein2013, Vogelsberger2013} and empirically derived mass loading factors are indicated by the colored curves \citep[i.e. This work;][]{Peeples2011, Lilly2013}\footnote{The mass loading factor derived by \citet{Lilly2013} cannot be parameterized by a power law in virial velocity (Equation \ref{eq:zeta}) and therefore the model parameters do not appear in Table 1.}. The functions are summarized in Table 1. Based on our empirical constraints from the oxygen census, we have derived the effective metal loading factor and not the mass loading factor in Section 3. We plot the effective metal loading factor from this work in Figure \ref{fig:eta}A under the assumption that $\gamma_w = 1$ and $\eta_w \gamma_w >> \eta_{acc} \gamma_{acc}$ in which case $\eta_w = \zeta$ in Equation \ref{eqn:best_fit_metal_loading}. 

Figure \ref{fig:eta}A demonstrates that there is a large discrepancy between empirically derived mass loading factors and those currently applied in theoretical models. The discrepancy results from the different observational constraints and set of assumptions adopted in theoretically and empirically motivated studies of mass loss. Theoretical mass loading factors are primarily constrained by the observed baryonic mass content of dark matter halos. Because galaxies are baryon deficient relative to the cosmological baryon fraction, massive outflows are required. However, interaction and coupling between the multi-phase outflowing material with pristine, infalling gas from the IGM is currently not well understood and this interaction may suppress initial gas accretion on galaxies \citep{vandeVoort2011}. In contrast, empirically derived mass loading factors attempt to produce the observed metal content of galaxies (e.g. the MZ relation). The primary assumption being that the reaccretion of metals ejected from the galaxy ISM is negligible. However, this is taken as a matter of convenience since efficiency of reaccretion is not observationally constrained.

In light of the observational uncertainties and varying set of assumptions, several possibilities exist for alleviating the discrepancy between theoretically and empirically derived mass loading factors : 
\begin{enumerate}
\item Some form of feedback suppresses initial gas accretion onto dark matter halos and/or galaxies \citep[see][]{vandeVoort2011}.  In this case, massive outflows are not required in order to reproduce the baryonic mass distribution of galaxies and theoretical mass loading factors in Figure \ref{fig:eta}A are unrealistically large. 

\item The metallicity of outflowing gas is extremely low (i.e. $\gamma_w << 1$).  While a large amounts of gas are expelled from galaxies, only a small fraction of the metals in the ISM are carried by the outflowing material. If we assume metal reaccretion is negligible (i.e. $\eta_{acc} \gamma_{acc} << 1$),  then we can solve Equation \ref{eq:zeta} for the metallicity outflowing gas such that $\gamma_w = \zeta/\eta_w$.  In Figure \ref{fig:eta}B we plot this for the various mass loading factors. \citet{Vogelsberger2013}, for example, find that they need $\gamma_w<1$ in order to simultaneously reproduce the galaxy stellar mass function and the MZ relation of local galaxies. The wind metallicity required in their simulations such that they are consistent with observed MZ relation of local galaxies (assuming $\eta_{acc} \gamma_{acc} << 1$) is shown by the dotted curve in Figure \ref{fig:eta}B.

\item Metal reaccretion is very efficient ($\gamma_{acc}\eta_{acc} >>  1$). Massive outflows may expel a large mass of metals but metals are able to preferentially accrete back onto galaxies \citep[e.g.][]{Dave2011b}. Equation \ref{eq:zeta} can be solved for the metal reaccretion efficiency such that $\eta_{acc} \gamma_{acc} = \gamma_w \gamma_w- \zeta$. We plot the metal reaccretion efficiency in Figure \ref{fig:eta}C sssuming that outflowing gas has the metallicity of the ISM ($\gamma_w = 1$).

\end{enumerate}
Most likely a combination of factors will likely contribute to resolving the discrepancy. Quantifying the properties of gas flows in Equation \ref{eq:int_oxy_def} is a key step for understanding the formation and evolution of galaxies. The effective metal loading factor derived in Section 3 provides straightforward way to apply empirical constraints in future theoretical models investigating mass loss and the apparent discrepancy.

\section{Summary and Conclusions}

We provide an important new parameterization for the empirical determination of the mass of oxygen expelled from star-forming galaxies first presented in \citet{Zahid2012b}. We present these constraints in the form of an effective metal loading factor which combines the uncertainties of outflows and inflows into a single factor. The effective metal loading factor is fit to the empirical determination of the mass of oxygen expelled from star-forming galaxies. This formulation provides joint constraints on the composition and magnitude of galaxy scale outflows. These constraints are parameterized as a function of halo virial velocity and thus may be straightforwardly implemented in theoretical models of galaxy evolution. 

We show that there is large discrepancy between theoretical and empirical constraints for mass loss in galaxies. The discrepancy is due to a differing set of observational constraints and assumptions adopted in the empirical and theoretical approaches, respectively. Possibilities for resolving the discrepancy are: 1) initial gas accretion onto dark matter halos is significantly suppressed; 2) outflows are extremely metal-poor \citep[see also][]{Vogelsberger2013, Torrey2013}; 3) metal reaccretion is extremely efficient; 4) all of the above. Our formulation of the empirical constraints on composition and magnitude of galaxy scale outflows may be implemented in theoretical models in order to explore these issues.

Our estimates of the mass of oxygen expelled from star-forming galaxies over their lives scales with galaxy stellar mass \citep{Zahid2012b}. This may be a result of outflows driven by dust which operate more efficiently in massive galaxies \citep{Zahid2013c}. Recent observations of an empirical upper limit for the gas-phase abundance in galaxies that is independent of redshift suggests that wind composition does not evolve significantly \citep{Zahid2013b}. Therefore, the derived effective metal loading factor may be applicable to the galaxy population over much of cosmic time. We anticipate the empirical estimates we derive for the effective metal loading factor will be useful for constraining outflows in cosmological simulations.  Future observations characterizing the physical properties of galaxy winds will provide important information for removing degeneracies between outflow composition and metal reaccretion.

\acknowledgments
HJZ thanks Margaret Geller and the Smithsonian Astrophysical Observatory for the gracious hospitality enjoyed during the period when much of this work was produced. We also like to thank Peter Behroozi for help in implementing the stellar to halo mass conversion used in this work.

\bibliographystyle{spr-mp-nameyear-cnd}
\bibliography{manuscript}

 \end{document}